\documentclass[final,1p,times]{elsarticle} 
\usepackage{graphicx} 
\usepackage{amssymb} 
\usepackage{amsthm} 
\usepackage{lineno} 

\journal{Nuclear Physics A} 
\begin{document} 

\begin{frontmatter} 


\title{Fixed scale approach to the equation of state on the lattice}

\author{K. Kanaya$^{a}$,  T. Umeda$^{b}$, S. Aoki$^{a,c}$, 
S. Ejiri$^{d}$, T. Hatsuda$^{e}$, \\
N. Ishii$^{e}$, 
Y. Maezawa$^{f}$ and 
H. Ohno$^{a}$  (WHOT-QCD collaboration)}

\address[a]{Graduate School of Pure and Applied Sciences, University of Tsukuba, Tsukuba, Ibaraki 305-8571, Japan}
\address[b]{Graduate School of Education, Hiroshima University, Hiroshima 739-8524, Japan}
\address[c]{RIKEN BNL Research Center, Brookhaven National Laboratory, Upton, New York 11973, USA}
\address[d]{Physics Department, Brookhaven National Laboratory, Upton, New York 11973, USA}
\address[e]{Department of Physics, The University of Tokyo, Tokyo 113-0033, Japan}
\address[f]{En'yo Radiation Laboratory, Nishina Accelerator Research Center, RIKEN, Wako 351-0198, Japan}

\begin{abstract} 
We propose a fixed scale approach to calculate the equation of state (EOS) in lattice QCD.
In this approach, the temperature $T$ is varied by $N_t$ at fixed lattice spacings. 
This enables us to reduce $T=0$ simulations 
which are required to provide basic data in finite temperature studies but are quite expensive in the conventional fixed-$N_t$ approach.
Since the conventional integral method to obtain the pressure is inapplicable at fixed scale, 
we introduce a new method, "T-integration method", 
to calculate pressure non-perturbatively.
We test the fixed scale approach armed with the T-integral method in quenched QCD on isotropic and
anisotropic lattices. 
Our method is found to be powerful to obtain reliable results for the
equation of state, especially at intermediate and low temperatures.
Reduction of the computational cost of $T=0$ simulations is indispensable to study EOS in QCD with dynamical quarks.
The status of our study in $N_f=2+1$ QCD with improved Wilson quarks is also reported.
\end{abstract} 

\end{frontmatter} 



\section{The method}

The equation of state (EOS) of hot QCD is an essential input 
to understand thermal properties of the quark matter in early Universe and in relativistic heavy ion collisions.
The lattice QCD is the only systematic method to calculate EOS 
for wide range of $T$ across the phase transition.
The temperature on the lattice is given by the inverse of the temporal lattice extent:
$T=(N_t a)^{-1}$ with $N_t$ the lattice size in the euclidian temporal direction and $a$ the lattice spacing.
Conventionally, finite temperature lattice simulations are performed in the fixed-$N_t$ scheme, 
in which $T$ is varied by changing $a$ (or equivalently the lattice gauge coupling $\beta = 6/g^{2}$) at a fixed $N_t$. 
Because this method requires a large amount of computational resources, 
most studies have been done with staggered-type lattice quarks which are less demanding among lattice quarks but whose fundamental properties such as locality and universality are not well established.
Therefore, to evaluate the effects of lattice artifacts, it is indispensable to carry out studies using theoretically sound lattice quarks, such as the Wilson-type quarks.
For this purpose, we developed the fixed scale approach with the T-integral method which can reduce numerical costs \cite{Tintegral}.

To study EOS, it is convenient to first calculate the trace anomaly $\epsilon-3p$ where $\epsilon$ is the energy density and $p$ is the pressure.
For the case of quenched QCD, 
\begin{equation}
\frac{\epsilon-3p}{T^4} = \frac{N_t^3}{N_s^3} \, a\frac{d\beta}{da} 
\left\{
\left\langle \frac{\partial S}{\partial\beta} \right\rangle
- \left\langle \frac{\partial S}{\partial\beta} \right\rangle_{T=0}
\right\}
\end{equation}
where 
$a\frac{d\beta}{da} $ is the lattice beta-function whose non-perturbative values are relatively easy to evaluate.
The subtraction of the zero-temperature counterpart is required for renormalization.
To obtain a non-perturbative estimate of $p$,  the integral method is usually adopted:
Using the thermodynamic relation $p = (T/V) \ln Z$ valid in the large volume limit, 
\begin{equation}
p = \frac{T}{V} \int^{\beta}_{\beta_0} \! d\beta \, \frac{1}{Z}
\frac{\partial Z}{\partial \beta} 
= -\frac{T}{V} \int^{\beta}_{\beta_0} \! d\beta 
\left\{
\left\langle \frac{\partial S}{\partial \beta} \right\rangle 
- \left\langle \frac{\partial S}{\partial\beta} \right\rangle_{T=0}
\right\}
\end{equation}
with $p(\beta_0) \approx 0$.
In this conventional method, major part of the computational cost
is devoted to zero temperature simulations to set the lattice scale, 
to carry out zero-temperature subtractions, 
and to determine the non-perturbative beta functions, at each simulation point. 
In the full QCD, lines of constant physics (LCP's) should be determined too at $T=0$.

In this report, we push an alternative approach where
temperature is varied by $N_t$ with other parameters fixed, i.e.\ at a fixed point in the coupling parameter space.
While this enables us to commonly use one $T=0$ simulation for all temperatures, the conventional integral method in the coupling parameter space is inapplicable.
Therefore, we propose a new method, ``the $T$-integral method'', in \cite{Tintegral} to calculate the pressure non-perturbatively:
Using a thermodynamic relation valid at vanishing chemical potential, we obtain
\begin{eqnarray}
T \frac{\partial}{\partial T} \left( \frac{p}{T^4} \right) =
\frac{\epsilon-3p}{T^4}
\hspace{6mm}\Longrightarrow\hspace{6mm}
\frac{p}{T^4} = \int^{T}_{T_0} dT \, \frac{\epsilon - 3p}{T^5}
\label{eq:Tintegral}
\end{eqnarray}
with $p(T_0) \approx 0$.

Since the coupling parameters are common to all temperatures,
$T=0$ subtractions can be done by a common zero temperature simulation, 
the condition to follow the LCP is obviously satisfied, and 
the lattice scale as well as beta functions are required only at
the simulation point.  
Then, the computational cost needed for $T=0$ simulations is reduced largely.
We may even borrow results of existing high precision spectrum studies at $T=0$
which are public e.g.\ on the International Lattice Data Grid (ILDG) \cite{ILDG}.
Because the lattice spacings in spectrum studies are usually
smaller than those used in conventional fixed-$N_t$ studies around the 
critical temperature $T_c$, for thermodynamic quantities around $T_c$, 
we can largely reduce the lattice artifacts due to large $a$ and/or small $N_t$ 
over the conventional approach. 
On the other hand, as $T$ increases, $N_t$ in our approach becomes small and hence the
lattice artifact increases.  
Therefore, our approach is not suitable for studying the high $T$ limit. 
Note that the merits and demerits of our method are complement to the conventional method. 
Our merits around $T_c$ may be a good news for phenomenological applications, since
temperatures relevant at RHIC and LHC will be at most up to a few times $T_c$. 

\section{Test in quenched QCD}

\begin{table}[tb]
\begin{center}
\begin{tabular}{c|cccccccc}
\hline
set & $\beta$ & $\xi$ & $N_s$ & $N_t$ &$r_0/a_s$ & $a_s$[fm] &
$L$[fm] & $a(dg^{-2}/da)$ \\
\hline
i1 & 6.0 & 1 & 16 & 3-10 & 5.35($^{+2}_{-3}$) & 0.093 & 1.5 & -0.098172 \\
i2 & 6.0 & 1 & 24 & 3-10 &5.35($^{+2}_{-3}$) & 0.093 & 2.2 & -0.098172 \\
i3 & 6.2 & 1 & 22 & 4-13 & 7.37(3) & 0.068 & 1.5 & -0.112127 \\
\hline
a2 & 6.1 & 4 & 20 & 8-34 & 5.140(32) & 0.097 & 1.9 & -0.10704 \\
\hline
\end{tabular}
\end{center}
\vspace*{-4mm}
\caption{Simulation parameters on isotropic and anisotropic lattices \cite{Tintegral}.
The beta function is taken from \cite{Boyd:1996bx}.
Corresponding $T=0$ simulations are done on $N_t=20\, \xi$ lattices.  
} 
\label{tab:para1}
\end{table}

We test the method in quenched QCD using the standard plaquette gauge action  \cite{Tintegral}. 
Simulation parameters are summarized in Table \ref{tab:para1}.
The i1, i2 and i3 lattices are isotropic, while the a3 lattice is anisotropic with $\xi \equiv a_s/a_t= 4$.
The temperature ranges cover $T \sim 200$--700 MeV.

\begin{figure}[tb]
\centering
\includegraphics[width=0.405\textwidth]{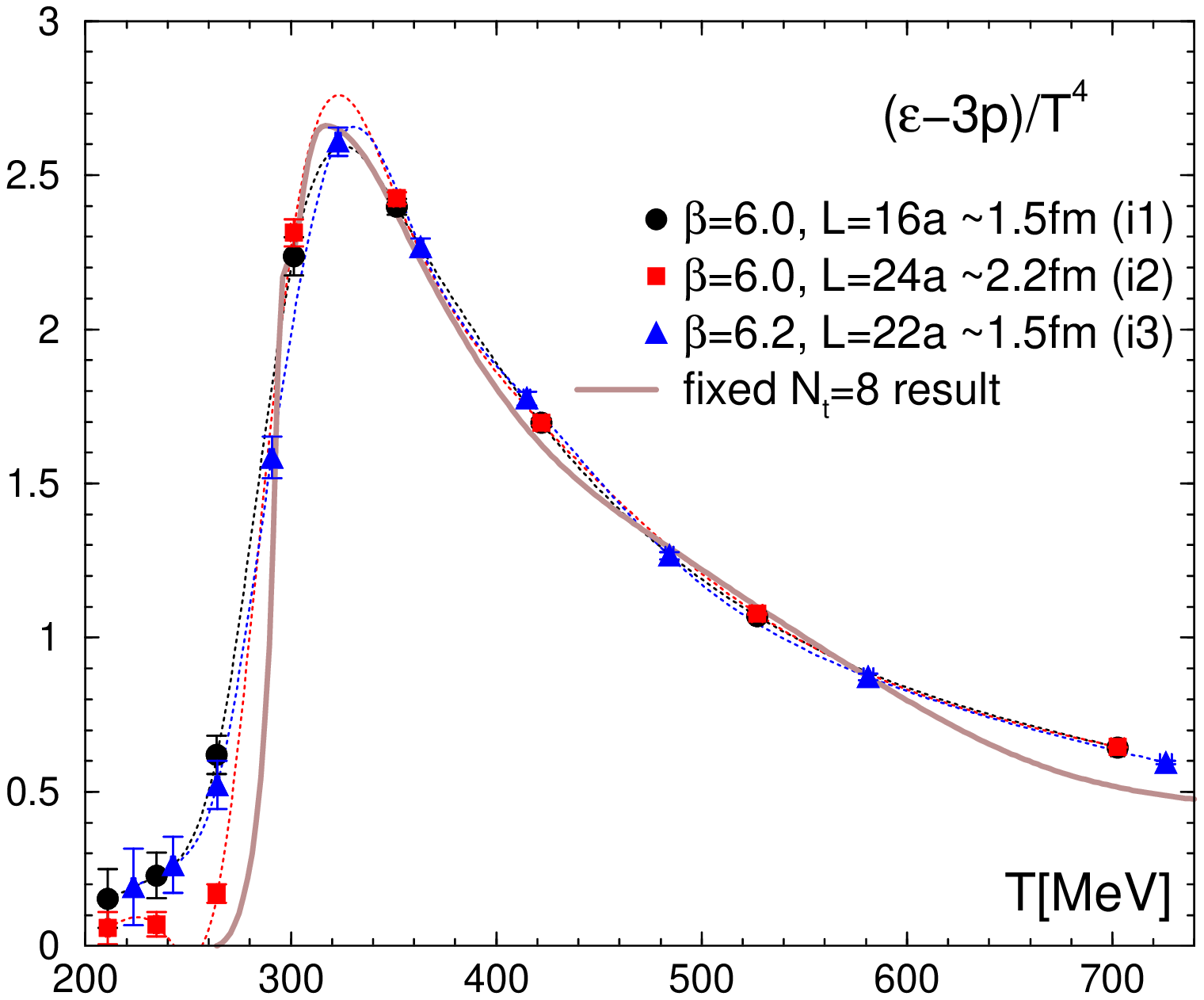}
\hspace{7mm}
\includegraphics[width=0.39\textwidth]{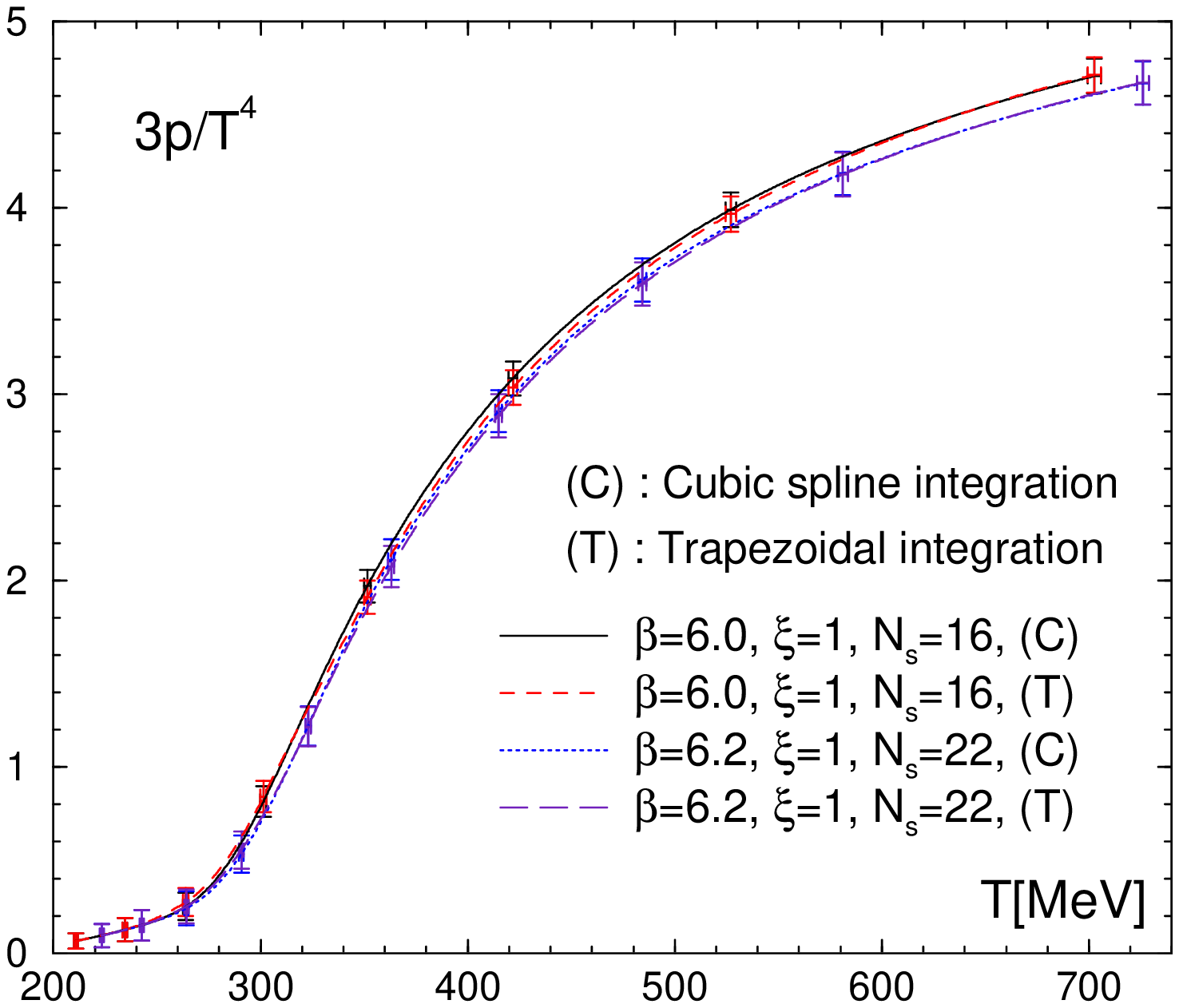}
\vspace*{-1.5mm}
\caption[]{EOS of quenched QCD on isotropic lattices by the T-integral method.
Left: trace anomaly.
Right: pressure.}
\label{fig1}
\end{figure}

The trace anomaly obtained on isotropic lattices are summarized in the left panel of Fig.\ref{fig1}.
The shaded line represents the result of the conventional fixed-$N_t$ method obtained on a quite large lattice of $N_t=8$ and the spatial lattice size $N_s=32$ (about 2.7 fm around $T_c\sim 290$ MeV) \cite{Boyd:1996bx}.
Comparing the results on i1 and i3 lattices, we find that the lattice cutoff effects are quite small on these lattices.
On the other hand, the i2 and $N_t=8$ lattices show small deviations near $T_c$. 
These deviations may be explained by the physical finite size effect expected in the critical region.
Off the critical region, all results agree well with each other.

\begin{figure}[tb]
\centering
\includegraphics[width=0.415\textwidth]{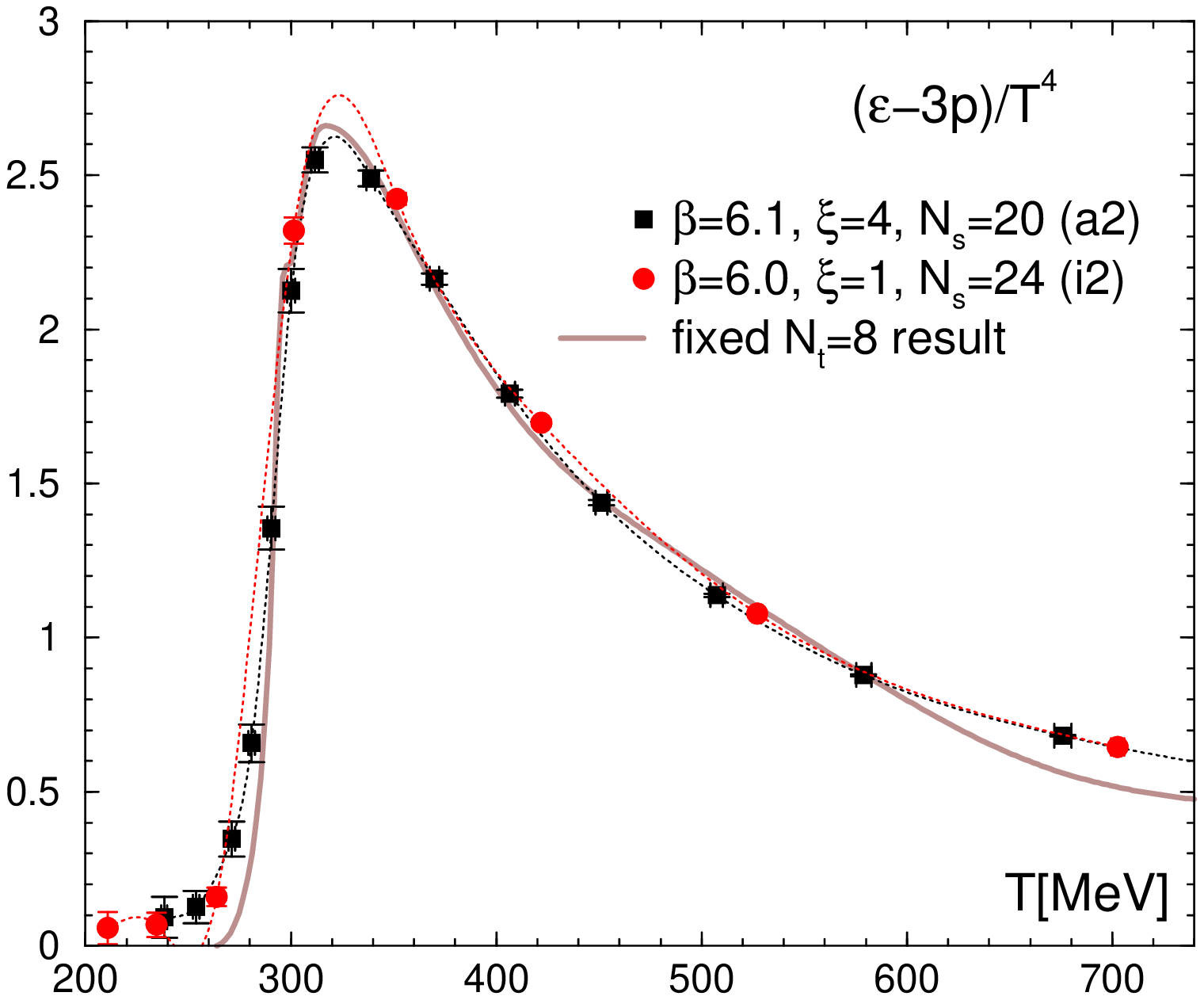}
\hspace{7mm}
\includegraphics[width=0.40\textwidth]{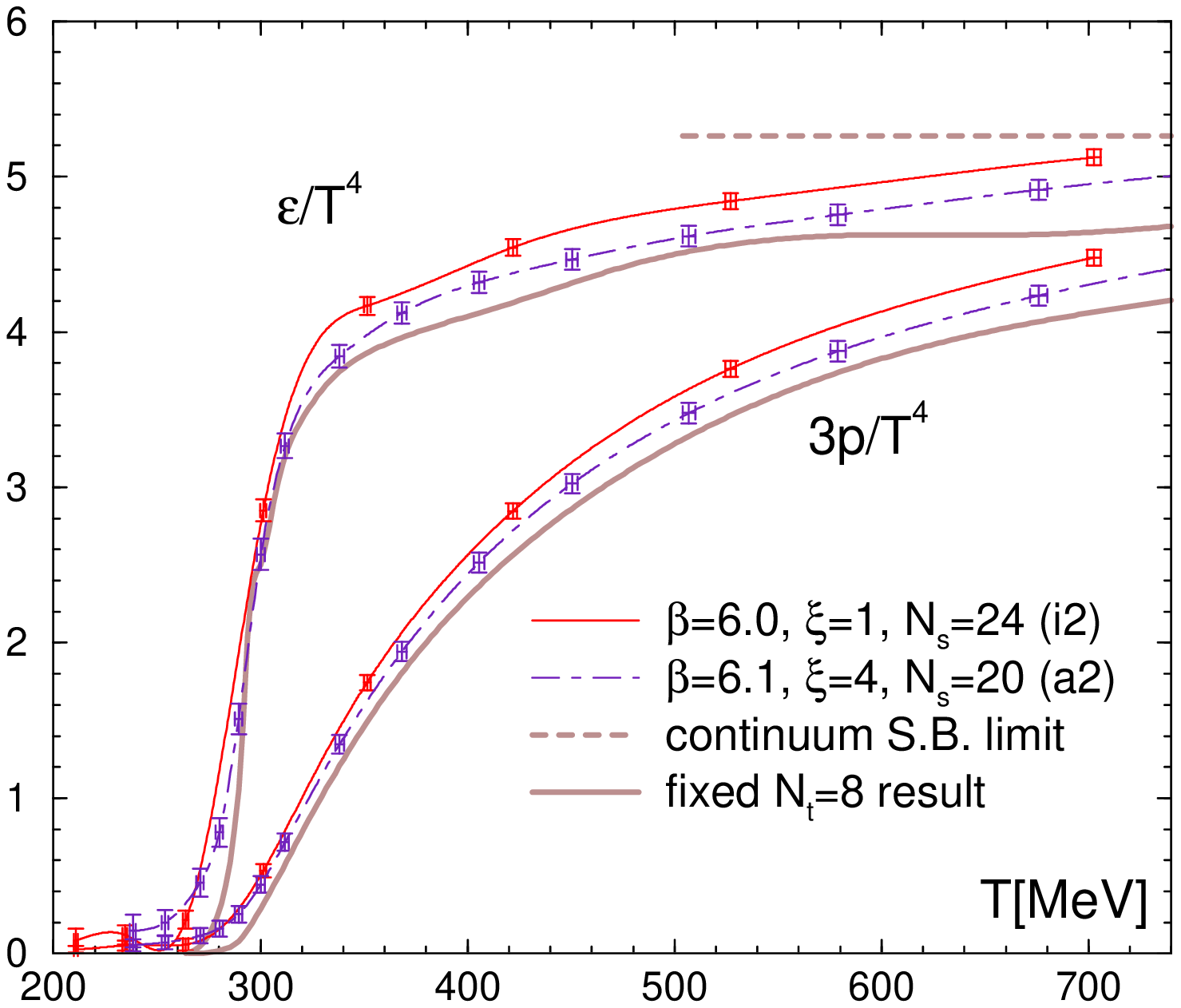}
\vspace*{-1.5mm}
\caption[]{EOS on isotropic i2 and anisotropic a2 lattices.
Left: trace anomaly.
Right: energy density and pressure.}
\label{fig2}
\end{figure}

Dotted lines in the  left panel of Fig.\ref{fig1} are the natural cubic spline interpolations. 
We have sufficiently many data to smoothly interpolate.
Carrying out the numerical integration (\ref{eq:Tintegral}), we obtain the pressure shown in the right panel.
Error bars shown in the figures represent the statistical errors estimated by the jackknife method.
To estimate the systematic error due to the interpolation ansatz, we show the results of interpolations with the trapezoidal rule (labeled by ``T'') too.
We find that all the results agree well with each other within the statistical errors.

In the T-integral method, $T$ is restricted to discrete values due to the discreteness of $N_t$.
We attempt to directly estimate systematic effects due to discreteness of $T$ by comparing the results with those obtained on an anisotropic lattice a2, which has about 4 times finer resolution in $T$ than the i2 lattice.
In Fig.\ref{fig2} left, we compare the trace anomaly on a2 and i2 lattices.
We find that the data points from the a2 lattice appear on the interpolation line from the i2 lattice,
while, due to the cruder resolution in $T$, 
the natural cubic spline interpolation for the i2 lattice slightly overshoots the data on the
a2 lattice around the peak. 
We note that the height of the peak on the a2 lattice is consistent to those of the fine i3 and $N_t=8$ lattices.
Carrying out the numerical integration, we obtain EOS shown in Fig.\ref{fig2} right.
We find that the results are well consistent with each other, suggesting that the systematic errors due to the discreteness of $T$ is at most about the statistical errors. 
See Ref.\cite{Tintegral} for more discussions.

\section{Outlook towards the EOS in QCD with dynamical $u$, $d$, $s$ quarks}

We have presented a new way to calculate EOS on the lattice.
Our method is complementary to the conventional method, and will have an advantage around the transition temperature.
We note that our method can be easily applied to finite densities using the conventional Taylor expansion method.

We are currently investigating EOS in $2+1$ flavor QCD with
non-perturbatively improved Wilson quarks,  
using the configurations by the CP-PACS/JLQCD Collaboration
\cite{CP-PACS-JLQCD}, which are public on the ILDG \cite{ILDG}. 
Among their simulation points, we have chosen the finest lattice ($a=0.07$ fm) with the lightest u and d quarks  ($m_\pi/m_\rho =0.63$).
Using the same coupling parameters, we are generating finite temperature configurations on $32^3\times N_t$ lattices 
with $N_t=4$, 6, $\cdots$, 16,
where the pseudo-critical temperature is expected to be around 14. 
Preliminary results for heavy quark free energies on these finite temperature configurations are presented by Maezawa in these proceedings \cite{Maezawa}, in which another good feature of the fixed scale approach is pointed out: 
We can study the thermal effects on the heavy quark free energies without artificially adjusting the constant term of the potential.
We hope that we present new results of EOS at the next QM symposium.
Our final goal is to study EOS in $2+1$ flavor QCD just at the physical quark mass point.
We are planning to extend the study using zero-temperature configurations 
being generated by the PACS-CS Collaboration
just at the physical point \cite{PACS-CS}. 


We thank the members of the CP-PACS and JLQCD Collaborations for providing us with the high-statistics $N_f=2+1$ QCD configurations.
This work is in part supported 
by Grants-in-Aid of the Japanese Ministry
of Education, Culture, Sports, Science and Technology, 
(Nos.17340066, 
18540253, 
19549001, 
20105001, 20105003, 
20340047, 
21340049  
). 
SE is supported by U.S.\ Department of Energy (DE-AC02-98CH10886). 
The quenched simulations have been performed on supercomputers 
at RCNP, Osaka University and YITP, Kyoto University. 
This work is in part supported also by the Large Scale Simulation Program of High Energy Accelerator Research Organization (KEK) Nos. 08-10 and 09-18.


\begin{thebibliography}{00} 
   

\bibitem{Tintegral} T. Umeda et al. (WHOT-QCD Collaboration), 
{\it Phys. Rev. D} {\bf 79} (2009) 051501.

\bibitem{Boyd:1996bx} 
G. Boyd {\it et al.}, {\it Nucl.\ Phys.\ B} {\bf 469} (1996) 419.

\bibitem{ILDG}
T. Yoshi\'e, 
{\it PoS} (LAT08) (2008) 042.

\bibitem{CP-PACS-JLQCD} 
T. Ishikawa et al. (CP-PACS and JLQCD Collaborations),
{\it Phys. Rev. D} {\bf 78} (2008) 011502.

\bibitem{Maezawa}
Y. Maezawa et al. (WHOT-QCD Collaboration), 
in these proceedings.

\bibitem{PACS-CS} 
Y. Kuramashi (PACS-CS Collaboration),
{\it PoS} (LAT08) (2008) 042.



\end{thebibliography}
\end{document}